\documentclass{desyproc}

\input paperdef

\newcommand{\sixooo}{600 \ifb}
\newcommand{\sixoooeff}{600 \ifb\,eff$\times2$}

\begin{document}
\title{Status of the Exclusive MSSM Higgs production at LHC after the Run I}

\author{{\slshape Marek Ta\v{s}evsk\'{y}}\\[1ex]
Institute of Physics of the Academy of Sciences of the Czech 
Republic, Na Slovance 2, \\
18221 Prague, Czech Republic}



\acronym{EDS'13} 

\maketitle

\begin{abstract}
We investigate the prospects for Central Exclusive Diffractive (CED)
production of MSSM Higgs bosons at the LHC using forward proton detectors (FPD)
proposed to be installed 220~m and 420~m from ATLAS and CMS detectors.
We summarize the situation after the first and very successful data taking 
period of the LHC. The discovery of a Higgs boson and results from searches
for additional MSSM Higgs bosons from the ATLAS and CMS, have recently led
to a proposal of new low-energy MSSM benchmark scenarios. The CED signal cross 
section for the process $H/h\!\to\!\bb$ and its backgrounds are estimated in
these new scenarios. We also comment on the experimental 
procedure if the proposed FPDs are to be used to measure the CED signal.
\end{abstract}

\section{Introduction}
A brief overview of the analysis is given here, while more details can be 
found in \cite{MT}.
The interest in the CED production of new particles is still
significant over the last decade (e.g. \cite{MT,KMRProsp,HarlandLang:2013jf} 
and references in \cite{MT}). 
The process is defined as $pp\rightarrow p\oplus\Phi\oplus p$
where all of the energy lost by the protons during the interaction
(a few per cent) goes into the production of the central system, $\Phi$. The 
final state therefore consists of a centrally produced system (e.g. dijet, 
heavy particle or Higgs boson) coming from a hard subprocess, two very forward 
protons and no other activity. The '$\oplus$' sign denotes the regions devoid 
of activity, often called rapidity gaps. A simultaneous detection of both 
forward protons and the central system opens up a window to a rich physics 
program covering not only exclusive but also a variety of QCD, Electroweak and
beyond Standard Model (BSM) processes (see e.g. 
\cite{KMRProsp,CMS-Totem,FP420,AFP,PPS}). Such measurements can
put constraints on the Higgs sector of Minimal Supersymmetric SM (MSSM) and 
other popular BSM scenarios \cite{KKMRext,diffH,CLP}. 

The attractivity of the CED production stems from a precise measurement of the
Higgs mass using FPDs, a possibility to measure its spin, parity and couplings 
to b-quarks using a few events and a good S/B ratio. These aspects
together with calculations of signal and background processes are in detail 
discussed in \cite{diffH}. Studies of the CED Higgs production contributed to 
the physics motivation for upgrade 
projects to install FPDs at 420~m \cite{FP420} and 220~m from the ATLAS (AFP 
project \cite{AFP}) and CMS (PPS project \cite{PPS}) detectors. At present, 
only 220~m devices are being considered in ATLAS and CMS. 

In MSSM~\cite{susy} the Higgs sector consists of five physical states. At the
lowest order the 
MSSM Higgs sector is $\cp$-conserving, containing two $\cp$-even bosons, the 
lighter $h$ and the heavier $H$, a $\cp$-odd boson, $A$, and the charged bosons
$H^\pm$. It can be specified in terms of the gauge couplings, the ratio of the 
two vacuum expectation values, $\tb \equiv v_2/v_1$, and the mass of the $A$
boson, $\MA$.  

Last year, the discovery of a new resonance with mass close to 125.5~GeV has 
been announced by ATLAS \cite{HdiscA} and CMS \cite{HdiscC}. 
Preliminary estimates of its spin-parity and couplings suggest that it behaves
like a SM Higgs boson. At the same time,
results from analyses searching for the MSSM signal at LHC have been 
published. Based on all these results i) seven new low-energy MSSM 
benchmark scenarios have been proposed \cite{newscenarios} that are 
compatible over large parts of the ($\MA$, $\tb$) parameter plane 
with the mass and production rates of the observed Higgs boson signal at 
125.5~GeV, and ii) the most recent LHC exclusion regions have been evaluated 
using the latest version of the program {\tt HiggsBounds} \cite{higgsbounds}. 
The aim of this analysis is to
investigate the CED Higgs boson production in these new benchmark scenarios 
taking into account the recent LHC exclusion regions and the region of the 
allowed Higgs mass. 

\section{Prospects in new benchmark MSSM scenarios}
\label{sec:bench}
\begin{figure}[htb!]
\vspace{-0.7cm}
\centerline{\includegraphics[width=10cm,height=6.2cm]
                {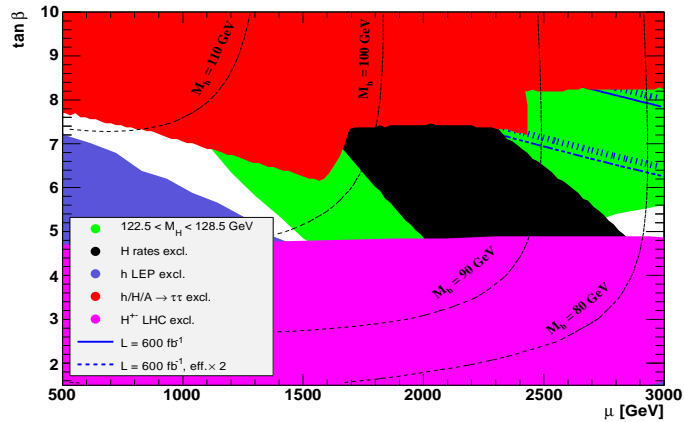}}
\vspace{-0.5cm}
\caption{Contours of $3\,\si$ statistical significance (solid blue lines) 
for the $h \to b \bar b$ channel in CED production at $\sqrt s = 14$~TeV in the 
($\mu$, $\tb$) plane of the MSSM within the Low-MH benchmark 
scenario. The values of the mass of the light $\cp$-even Higgs boson, $\Mh$, are
indicated by dashed (black) contour lines. The dark shaded (blue) region 
corresponds to the parameter region that is excluded by the LEP MSSM Higgs 
searches, the lighter shaded (red), the lighter shaded (pink) and black areas 
are excluded by LHC MSSM Higgs searches in the analyses of h/H/A $\rightarrow 
\tau\tau$, charged Higgs and Higgs rates, respectively. The light shaded 
(green) area corresponds to the allowed mass region $122.5 < \MH < 128.5$~GeV. 
}
\label{fig:LowMH}
\vspace{-0.2cm}
\end{figure}

\vspace*{-0.2cm}
The SM cross section used~\cite{bbhatnnlo} for the normalization within 
{\tt HiggsBounds} is evaluated using the MRST2002 NNLO PDFs. 
For each point in the parameter space we evaluate the relevant Higgs
production cross section times the Higgs branching ratio for the $\bb$ decay in
MSSM (BR($h\!\to\!\bb$)). The values of $\MH$, BRs and effective couplings in 
MSSM are calculated with the program \fh~\cite{feynhiggs}. The resulting 
theoretical cross section is multiplied by the experimental efficiencies 
as described in \cite{diffH}. We also show the parameter regions excluded 
by the LEP and LHC Higgs-boson searches as obtained with 
{\tt HiggsBounds}~\cite{higgsbounds} and so called region of allowed Higgs 
masses, i.e. $\MH = 125.5 \pm 3$~GeV (by the light gray (green)). The total
uncertainty of 3~GeV represents a combination of the experimental 
($\!\sim\!\!0.6$~GeV) and theoretical uncertainty from unknown higher-order 
corrections in MSSM. The prospects for observing the neutral $\cp$-even Higgs 
bosons in CED within the new MSSM benchmark scenarios are discussed in detail 
in \cite{MT}. In summary: available cross-sections in all scenarios are
too small (smaller than 0.02~fb) to be considered seriously for further studies
with the exception of the Low-MH scenario for which the contours of $3\,\si$ 
statistical significances are shown in Fig.~\ref{fig:LowMH} for the 420+220 FPD 
configuration and for two assumed effective luminosities, \sixooo\ and 
\sixoooeff.
The region of interest is the area of allowed Higgs masses that is not overlaid
by the LHC exclusion region. 
The highest achievable significances are located in the same corner of the 
green band as the highest S/B ratios (see \cite{MT}). 
We conclude that if MSSM should be realized as in the Low-MH scenario, i.e. the
heavy Higgs at mass of 125.5~GeV and the lighter one in the range 
80--90~GeV, then FPD projects could be very helpful to ATLAS and CMS in 
searches for such a low-mass object. We stress that due to the mass acceptance 
this light Higgs boson can only be seen with stations at 420~m.
 
A few notes about experimental issues: i) the total integrated luminosity 
needed to observe the light Higgs boson produced in CED with mass around 
80--90~GeV is of the order of 1000~fb$^{-1}$ meaning that data from both 
the AFP and PPS would have to be combined, ii) putting the AFP stations 
at 420~m into the L1 ATLAS trigger scheme is currently impossible 
due to a short L1 latency,
iii) the total mass acceptance decreases and the mass resolution and b-tagging 
efficiency worsen with decreasing mass. 
However, improvements are expected in the reduction of background, e.g. in 
reducing the misidentification of gluon to be b-quark and resolution below 
10~ps in the fast timing detectors. We conclude that investigating 
the mass range 80--90~GeV with FPDs at LHC is 
more challenging than that around 120~GeV. 

\section*{Acknowledgments}
The work was supported by the project LG13009 of the Ministry of Education of 
the Czech republic. Big thanks to Sven Heinemeyer and Valery Khoze who 
participated at the early stage of this analysis for their encouragement and 
assistance.  


\vspace*{-0.1cm}
\begin{footnotesize}

\end{footnotesize}

\end{document}